\documentstyle[pre,aps,multicol]{revtex}

\begin{document}
\draft
\title{{\bf Influence of Long-range Interactions on the Critical Behavior of
Systems with negative Fisher-Exponent}}
\author{{\sc H.K.\ Janssen}}
\address{Institut f\"{u}r Theoretische Physik III, Heinrich-Heine-Universit\"{a}t,\\
40225 D\"{u}sseldorf, Germany }
\date{\today}
\maketitle

\begin{abstract}
The influence of long-range interactions decaying in $d$ dimensions
as $ 1/R^{d+\sigma }$ on the critical behavior of systems with
Fisher's correlation-function exponent for short-range interactions
$\eta _{SR}<0$, is re-examined. Such systems, typically described
by $\Phi ^{3}$-field theories, are e.g.\ the Potts-model in the
percolation-limit, the Edwards-Anderson spin-glass, and the
Yang-Lee edge singularity. In contrast to preceding studies, it is
shown by means of Wilson's momentum-shell renormalization-group
recursion relations that the long-range interaction dominates as
long as $\sigma <2-\eta _{SR}$. Exponents change {\em continuously
}to their short-range values at the boundary of this region.
\end{abstract}


\begin{multicols}{2}
\narrowtext

Fifteen years ago there was some debate about the influence of
long-range interactions (decaying as $J\left( R\right) \sim
1/R^{d+\sigma }$ in $d$ dimensions) on the critical exponents of
systems that show a negative Fisher-exponent $\eta _{SR}$ if the
long-range forces are absent \cite {ChS83,ThG84}. Finally it was
claimed that the long-range interactions leading to a
Fisher-exponent $\eta _{LR}=2-\sigma $ dominate for all $\sigma
<2$, and, by reason of an instability, the exponents change
discontinuously to their short-range values at $\sigma =2$. The
assumption that long-range interactions decaying with $\sigma >2$
are generally equivalent to purely short-range interactions seems
to be the accepted lore. Indeed in the Gaussian part of an
effective Landau-Ginzburg-Wilson Hamiltonian in momentum space
${\cal H}_{0}=\frac{1}{2}\int_{q}\left( r+j_{2}q^{2}+j_{\sigma
}q^{\sigma }+\ldots \right) s_{q}s_{-q}$ ($s_{q}$ is the Fourier
transform of an order parameter fluctuation and $\int_{q}\ldots
=\int d^{d}q\ldots $) where as usual $j_{2}$ stems from the
short-range part of the interaction (which also contributes to $r$)
and $j_{\sigma }$ from the long-range one, $ j_{\sigma }$ is
naively irrelevant in comparison to $j_{2}$ at long-wave lengths if
$\sigma >2$. Of course, this is an incorrect argumentation below
the upper critical dimension where the nonquadratic higher-order
terms of the Hamiltonian play a dominant role. The relevance of the
long-range term $
\propto q^{\sigma }$ (which does not renormalize because of its
nonanalyticity in $q$) has to be determined by comparison with the scaling
behavior of the full inverse correlation function $\Gamma _{2}\left(
q\right) _{SR}\propto q^{2-\eta _{SR}}$ that corresponds to the nontrivial
stable fixed point solution of a properly chosen renormalization group
transformation.

In the present letter I show that also in the case $\eta _{SR}<0$
the long-range interaction dominates as long as $2-\sigma =\eta
_{LR}>\eta _{SR}$
. At $\eta _{LR}=\eta _{SR}$ the exponents change {\em continuously} to
their short-range values that hold everywhere for $\sigma >2-\eta _{SR}$.
This behavior is well known for models with $\eta _{SR}>0$ \cite
{FMN72,Sa73,Ah76,Ya80,HN89} but is seemingly nonaccepted as yet for $\eta
_{SR}<0$. The incorrect result of \cite{ChS83,ThG84} with respect to the
crossover arises from a renormalization group which is not appropriate for
that case because it mixes in a redundant operator in the terminology of
Wegner \cite{We76}. I show that in general the limit $\sigma \rightarrow 2$
produces a contribution $\sim q^{2}\ln q$ to the Gaussian part of the
Hamiltonian that is a relevant perturbation here. Thus the critical
exponents are different for the cases with or without such a perturbation
that is therefore responsible for the apparent discontinuity of exponents.

Systems with a negative Fisher exponent $\eta _{SR}$ are typically described
by critical $\Phi ^{3}$-field theoretic models \cite{AKK80} with an upper
critical dimension $d_{c}=6$ such as the Potts model in the percolation
limit \cite{HLHD75,Am76,PL76}, the Yang-Lee singularity model \cite
{Fi78,BJ81}, and the Edwards-Anderson spin-glass model in the replica
formalism \cite{EA75,HLC76}. Besides these equilibrium models,
nonequilibrium models of $\Phi ^{3}$-type are given by epidemic processes
which lead to percolation clusters. Here long-range interactions creep in if
the disease spreads by Levy-flights \cite{Gr86,Ja98}. Treating these
nonequilibrium models I became aware of the earlier work \cite{ChS83,ThG84}
on long-range interactions in $\Phi ^{3}$-models.

In this letter I concentrate on the discussion of the Yang-Lee model as the
simplest among all others leading to similar recursion equations. To stick
close to the above mentioned work, I use the Wilson renormalization group
transformation based on the elimination of short-wave fluctuations and a
hard momentum cutoff normalized to $q_{c}=1$. I perform the renormalization
group to one-loop order and use $\varepsilon $-expansion where $\varepsilon
=6-d$. This is sufficient to produce the nontrivial crossover behavior,
since $\eta _{SR}=O\left( \varepsilon \right) $.

I write the Yang--Lee-Hamiltonian of the scalar field $s$ in the following
form
\begin{eqnarray}
{\cal H} &=&\int d^{d}x\,\left\{ \frac{1}{2}\left( \nabla s\right)
^{2}+ \frac{v}{4\alpha }\left[ \left( \nabla ^{1-\alpha }s\right)
^{2}-\left(
\nabla s\right) ^{2}\right] \right.   \nonumber \\
&&\qquad \qquad \left. +\frac{r}{2}s^{2}+\frac{ig}{6}s^{3}+ihs\right\} .
\label{1}
\end{eqnarray}
Here $\left( \nabla ^{1-\alpha }s\right) ^{2}$ is defined in momentum space
as $q^{2\left( 1-\alpha \right) }s_{q}s_{-q}$,and the momentum scale is
chosen such that $\left| q\right| \leq q_{c}=1$. I have written the
long-range exponent as $\sigma =2\left( 1-\alpha \right) $. Then $\alpha
=O\left( \varepsilon \right) $ in the crossover region. The Gaussian part of
${\cal H}$ with the derivatives of the field is positively definite
as long as $v\geq 0$ irrespective of the sign of $\alpha $, and
reads in momentum space $\frac{1}{2}q^{2}\left( 1-v\ln q\right)
s_{q}s_{-q}$ in the limit $
\alpha \rightarrow 0$. Thus also in this limit the model does not coincide
with its short-ranged counterpart unless $v=O\left( \alpha \right) $. The
unperturbed correlation function (the propagator of a diagrammatic
perturbation expansion) follows from (\ref{1}) as
\begin{equation}
G_{0}\left( q\right) =\left( q^{2}+\frac{v}{2\alpha }\left[
q^{-2\alpha }-1
\right] q^{2}+r\right) ^{-1}.  \label{2}
\end{equation}
Thus the scale of the fields is defined such that for $r=0$ the propagator
is $1$ at the cut-off momentum $q_{c}=1$. The propagator is positively
definite for all $q$ and $\alpha $ as it should be for stability.

The calculation of the momentum integrals that arise by eliminating
to one-loop order fluctuations $s_{q}$ which depend on momenta in
the interval $ b^{-1}<q\leq 1$ with $b\approx 1$ is standard and
does not present any technical difficulties. The coefficients of
the terms of different order in the Hamiltonian ${\cal H}$
(\ref{1}) change by the elimination procedure to
\begin{eqnarray}
&&q^{2}\left\{ 1+\frac{v}{2\alpha }\left( q^{-2\alpha }-1\right) \right\} +r
\nonumber \\
&\rightarrow &q^{2}\left\{ 1+\frac{v}{2\alpha }\left( q^{-2\alpha }-1\right)
-\frac{2uK\left( v,r\right) }{d\left( 1+r\right) ^{4}}\ln b\right\}
\nonumber \\
&&+\left\{ r+\frac{u}{\left( 1+r\right) ^{2}}\ln b\right\} +O\left(
q^{4},u^{2}\right)   \label{3}
\end{eqnarray}
and
\begin{eqnarray}
g &\rightarrow &g\left\{ 1-\frac{2u}{\left( 1+r\right) ^{3}}\ln b+O\left(
q^{2},u^{2}\right) \right\}   \label{4} \\
h &\rightarrow &h+\frac{u/g}{1+r}\ln b.  \label{4a}
\end{eqnarray}
Here, I have defined $u=\left( 4\pi \right) ^{-d/2}g^{2}/\Gamma \left(
d/2\right) $, and $K\left( v,r\right) $ is given by
\begin{eqnarray}
K\left( v,r\right)  &=&\frac{d-2}{4}+\frac{2+2\alpha -d}{8}v-\frac{v^{2}}{8}
\nonumber \\
&&-\frac{1-\alpha }{4}vr+\frac{2-v}{8}dr.  \label{5}
\end{eqnarray}
After elimination of the short-wavelength fluctuations an
appropriate rescaling has to be introduced as the last step of the
renormalization transformation. The goal is to choose a rescaling
$\zeta $ of the fields $ s^{<}\left( {\bf r}\right) =\zeta
s^{\prime }\left( b^{-1}{\bf r}\right) $ where $s^{<}\left( {\bf
r}\right) =\int_{q\leq b^{-1}}e^{i{\bf qr}}s_{{\bf q}}$ in such a
way that a renormalization group is constructed which leads to
fixed points and does not mix in {\em redundant dangerous
operators} \cite {We76}. The operator generated by a rescaling of
the fields without an elimination is such a redundant operator.
Thus the rescaling factor $\zeta $ must be chosen carefully. The
old (and working) definition for $\zeta $ follows from the
requirement to hold the propagator finite at the cutoff momentum to
exclude infrared singularities. Then such singularities cannot
arise in the full elimination procedure. Note that in the
long-range interaction problem normally one holds constant the
coefficient of the nonanalytic term $\sim q^{2\left( 1-\alpha
\right) }$. This may be the simplest possibility but leads to the
difficulty of a vanishing inverse propagator in the present case
\cite{ChS83,ThG84}. Therefore I define
\begin{equation}
\zeta ^{2}=b^{2-d-\gamma \left( u,v\right) }.  \label{6}
\end{equation}
Here the function $\gamma \left( u,v\right) $ follows from the
requirement $ G_{0}^{\prime }\left( q=1,r^{\prime }=0\right)
=G_{0}\left( q=1,r=0\right) =1 $ for the rescaled propagator. Now
by rescaling (\ref{3},\ref{4}), renormalized parameters are found
from
\begin{eqnarray}
&&q^{2}\left\{ 1+\frac{v^{\prime }}{2\alpha }\left( q^{-2\alpha }-1\right)
\right\}   \nonumber \\
&=&q^{2}b^{-\gamma }\left\{ 1+\frac{v}{2\alpha }\left( b^{2\alpha
}q^{-2\alpha }-1\right) -\frac{2uK\left( v,r\right) }{d\left(
1+r\right) ^{4} }\ln b\right\} ,  \label{7a}
\end{eqnarray}
and
\begin{eqnarray}
r^{\prime } &=&b^{2-\gamma }\left\{ r+\frac{u}{\left( 1+r\right)
^{2}}\ln b\right\} ,  \label{7} \\ u^{\prime } &=&ub^{6-d-3\gamma
}\left\{ 1-\frac{4u}{\left( 1+r\right) ^{3}}
\ln b\right\} ,  \label{8} \\
h^{\prime } &=&b^{\left( d+2-\gamma \right) /2}\left\{ h+\frac{u/g}{1+r}\ln
b\right\} .  \label{8a}
\end{eqnarray}
Taking (\ref{7a}) for $q=1$ yields the equation for $\gamma $:
\begin{equation}
1=b^{-\gamma }\left\{ 1+\frac{v}{2\alpha }\left( b^{2\alpha
}-1\right) -
\frac{2uK\left( v,r\right) }{d\left( 1+r\right) ^{4}}\ln b\right\} .
\label{9}
\end{equation}
Expanding $b=\exp l$ to first order in $l$ around $l=0$ one finds
\begin{eqnarray}
\gamma  &=&v-\frac{2uK\left( v,r\right) }{d\left( 1+r\right) ^{4}}  \nonumber
\\
&=&v-\frac{u}{3}\left( 1-\frac{2-\alpha }{4}v-\frac{v^{2}}{8}\right)
+O\left( \varepsilon ^{2}\right) .  \label{10}
\end{eqnarray}
In the last equation I have used $r=O\left( u\right) $, $u=O\left(
\varepsilon \right) $, and I have retained only terms linear in $\varepsilon
$. Note that the dependence on $v$ is exact to linear order in $u$.
For an infinitesimal transformation with $l=dl$ one gets the
renormalization group equations from the remaining parts of the
equations (\ref{7a},\ref{7},\ref{8},\ref{8a}) as
\begin{eqnarray}
\frac{du}{dl} &=&\left( \varepsilon -3\gamma -4u\right) u,  \label{11} \\
\frac{dv}{dl} &=&\left( 2\alpha -\gamma \right) v,  \label{12} \\
\frac{dr}{dl} &=&(2-\gamma )r+\frac{u}{\left( 1+r\right) ^{2}},  \label{13}
\\
\frac{dh}{dl} &=&\frac{d+2-\gamma }{2}h+\frac{u}{g\left( 1+r\right) }.
\label{14}
\end{eqnarray}
It follows from the nonanalyticity of the long--range term of the
Hamiltonian (\ref{1}) that $v$ does not acquire any contribution by the
elimination step and changes only under the rescaling. Thus the second
equation (\ref{12}) is exact, whereas the other three are correct only to
one-loop order.

The last two equations (\ref{13},\ref{14}) show in general fixed
points $ r_{\ast },h_{\ast }=O\left( u_{\ast }\right) $ for the
relevant parameters $r $ and $h$. Using these values the first two
equations (\ref{11},\ref{12}) in combination with $\gamma $
(\ref{10}) lead to four different fixed points for the coupling
constants $u$ and $v$. There are two Gaussian ones ($ u_{\ast
}=0$), namely a short-range fixed point with $v_{\ast }=0$, stable
for $\varepsilon <0$, $\alpha <0$ with $\eta :=\gamma _{\ast }=0$,
and a long-range fixed point with $v_{\ast }=\eta =2\alpha $,
stable for $
\varepsilon <6\alpha $, $\alpha >0$. Beside these trivial fixed points there
are two nontrivial ones with $u_{\ast }>0$. The well-known
\cite{Fi78,BJ81} short-range fixed point with $v_{\ast }=0$ follows
from (\ref{11}) as $ u_{\ast }=\varepsilon /3+O\left( \varepsilon
^{2}\right) $ and leads to $
\eta =\eta _{SR}=-\varepsilon /9+O\left( \varepsilon ^{2}\right) $. It is
stable for $\varepsilon >0$ as long as $2\alpha <\eta _{SR}$. But
if $ 2\alpha >\eta _{SR}$, it becomes unstable and the long-range
fixed point develops from (\ref{12}) with $v_{\ast }=2\alpha
+\left( \varepsilon
-6\alpha \right) /12+O\left( \varepsilon ^{2}\right) $, $u_{\ast }=\left(
\varepsilon -6\alpha \right) /4+O\left( \varepsilon ^{2}\right) $, and is
stable up to $\alpha =\varepsilon /6$, and $\eta =\eta _{LR}=2\alpha $. The
stability regions are shown in Fig.\ \ref{fig1}. In each case the
short-range behavior changes {\em continuously} to the long-range behavior
and vice versa at the line defined by $\eta _{SR}=\eta _{LR}$. For all fixed
points, the correlation length exponent $\nu $ follows from the linearized
equation (\ref{13}) for $r-r_{\ast }$. Here, in the case of the Yang-Lee
model, a Ward identity states $\beta =1$ for the order parameter exponent
\cite{AKK80,BJ81} and one obtains $\nu =2/\left( d-2+\eta \right) $ for both
nontrivial fixed points.

To get a picture of the renormalization flow of the coupling constants and
the movement of the fixed points for $\varepsilon >0$, I have rescaled the
variables as $x=u/\varepsilon $, $y=v/\varepsilon $, and introduced the
parameter $p=2\alpha /\varepsilon $. With a ``time'' $t=l/\varepsilon $ the
equations of motion are found as
\begin{eqnarray}
\dot{x} &=&\left( 1-3x-3y\right) x  \nonumber \\
\dot{y} &=&\left( p+\frac{x}{3}-y\right) y.  \label{15}
\end{eqnarray}
The flows and the fixed points are shown in Fig.\ \ref{fig2} for different
parameter values. Again one sees the {\em continuous }bifurcation of the
different fixed points corresponding to the {\em continuous} crossover of
short-range and long-range behavior.

In conclusion I have shown that an old result concerning the crossover
between long- and short-range interaction behavior in critical systems with
negative short-range Fisher exponent $\eta _{SR}$ is incorrect. As in the
case of a positive Fisher exponent, the behavior changes continuously at a
line defined by $\eta _{SR}=\eta _{LR}=2-\sigma $. The long-range
interactions dominate always as long as $\eta _{SR}<\eta _{LR}$.

\acknowledgments I thank S.\ Theiss for a critical reading of the
manuscript. This work has been supported in part by the SFB 237 (``Unordnung
und gro\ss e Fluktuationen'') of the Deutsche Forschungsgemeinschaft.

\begin{figure}[tbp]
\caption{Renormalization flow of the coupling constants $x=u/\protect
\varepsilon$, $y=v/\protect\varepsilon$ for different parameter values $p=2
\protect\alpha /\protect\varepsilon$ in the case of $\protect\varepsilon =
6-d > 0$ to one-loop order. The topology of the flow changes {\em
continuously} with $p$, and there exists always a stable long-range
($y > 0$ ) fixed point if $p > -1/9$.}
\label{fig1}
\end{figure}

\begin{figure}[tbp]
\caption{The stability regions of long- and short-range behavior. Nontrivial
short- (SR) and long-range (LR), as well as trivial Gaussian short-
(GSR) and Gaussian long-range (GLR) regions are shown. The behavior
changes {\em continuously} at the boundaries.}
\label{fig2}
\end{figure}

\end{multicols}

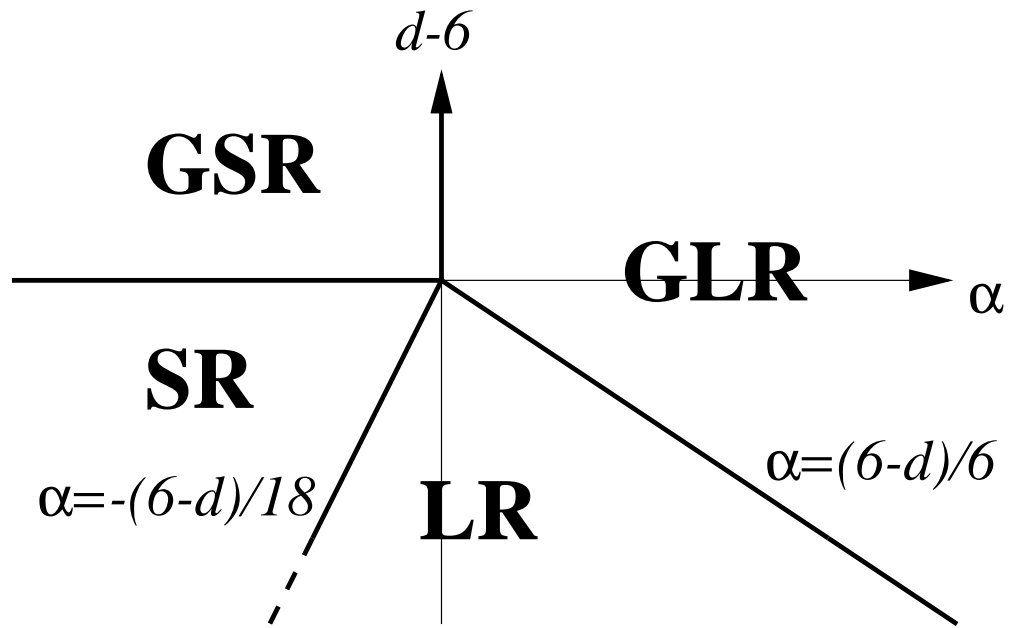
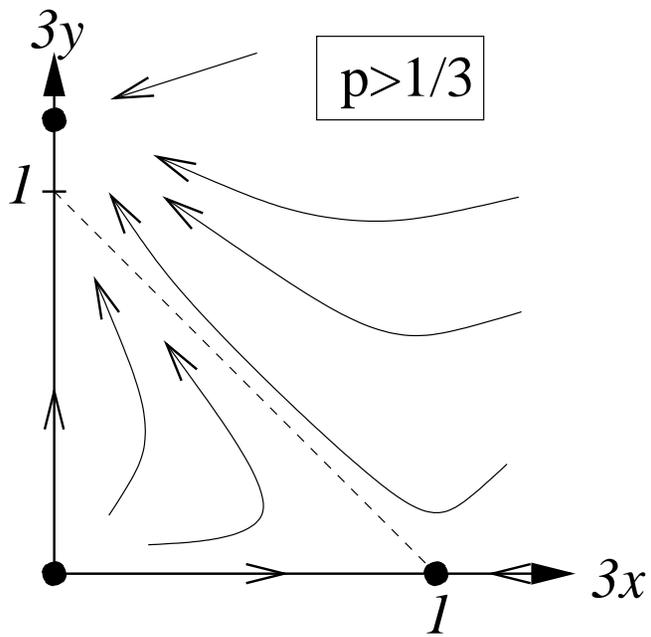
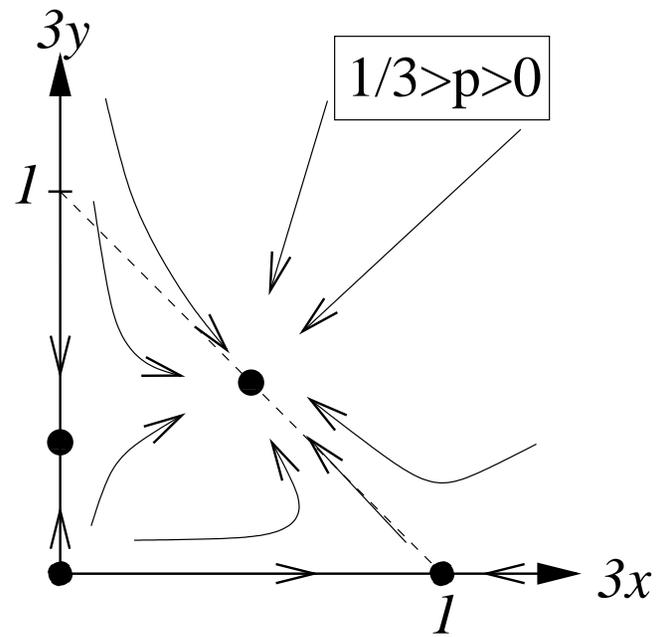
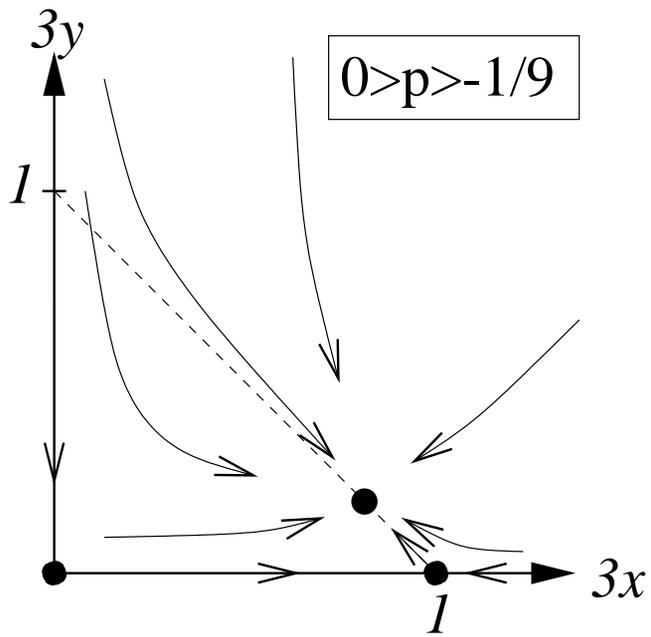
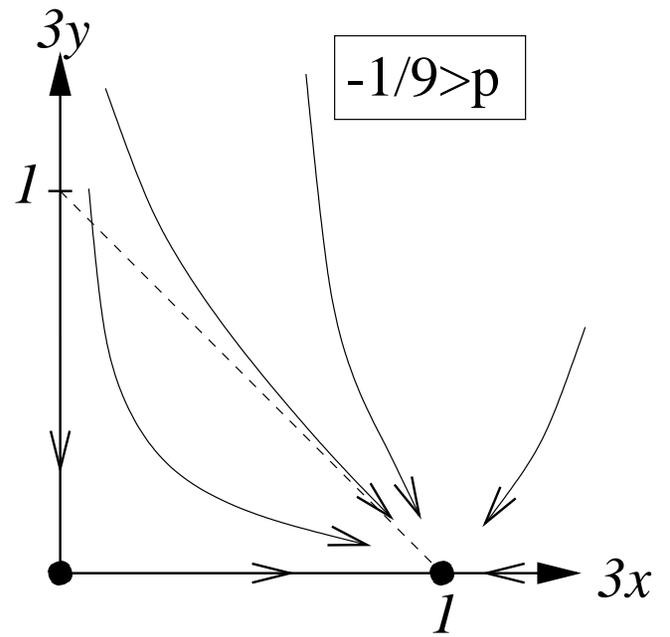
\end{document}